\documentclass[%
 aip,
 apl,
 amsmath,amssymb,
preprint,%
]{revtex4-1}
\pdfoutput=1
\paperwidth=\dimexpr \paperwidth + 6cm\relax
\oddsidemargin=\dimexpr\oddsidemargin + 3cm\relax
\evensidemargin=\dimexpr\evensidemargin + 3cm\relax
\marginparwidth=\dimexpr \marginparwidth + 2.85cm\relax

\usepackage{graphicx}
\usepackage{dcolumn}
\usepackage{bm}
\usepackage[font=footnotesize,caption=false]{subfig}

\usepackage[utf8]{inputenc}
\usepackage[T1]{fontenc}
\usepackage{mathptmx}
\usepackage{color}
\usepackage[english, textwidth=4cm, color=yellow, colorinlistoftodos]{todonotes}

\newcommand{\Voc}{V_{\mathrm{oc}}}

\newcommand{\nktq}{\frac{\eta k_BT}{q}}

\begin{document}

\title{Temperature and intensity dependence of the open-circuit voltage of InGaN/GaN multi-quantum well solar cells}

\author{Matthias Auf der Maur}
 \email{auf.der.maur@ing.uniroma2.it.}
\affiliation{ 
Department of Electronic Engineering, University of Rome Tor Vergata, 00133 Rome, Italy
}%
\author{Gilad Moses}%
\affiliation{%
Department of Solar Energy and Environmental Physics, Blaustein Institutes for
Desert Research, Ben-Gurion University of the Negev, Sede Boqer Campus, 8499000, Israel%
}%
\affiliation{%
Albert Katz School for Desert Studies, Blaustein Institutes for Desert Research,
Ben-Gurion University of the Negev, Sede Boqer Campus, 8499000, Israel
}%
\author{Jeffrey M. Gordon}%
\affiliation{%
Department of Solar Energy and Environmental Physics, Blaustein Institutes for
Desert Research, Ben-Gurion University of the Negev, Sede Boqer Campus, 8499000, Israel%
}%
\author{Xuanqi Huang}
\affiliation{%
School of Electrical, Computer and Energy Engineering, Arizona State University, Tempe, AZ 85287, USA
}%
\author{Yuji Zhao}
\affiliation{%
School of Electrical, Computer and Energy Engineering, Arizona State University, Tempe, AZ 85287, USA
}%
\author{Eugene A. Katz}
\affiliation{%
Department of Solar Energy and Environmental Physics, Blaustein Institutes for
Desert Research, Ben-Gurion University of the Negev, Sede Boqer Campus, 8499000, Israel%
}%

\date{\today}

\begin{abstract}
We have analyzed the temperature and intensity dependence of the open-circuit voltage of InGaN/GaN multi-quantum well solar cells up to 725 K and more than 1000 suns.
We show that the simple ABC model routinely used to analyze the measured quantum efficiency data of InGaN/GaN LEDs can accurately reproduce the temperature and intensity dependence of the measured open-circuit voltage if a temperature-dependent Shockley-Read-Hall lifetime is used and device heating is taken into account.
\end{abstract}

\maketitle

In recent years, InGaN/GaN multi-quantum well (MQW) structures have gained increasing interest for photovoltaic applications.\cite{Dahal2009,Bhuiyan2012,Williams2017,Park2018,Bai2018,Huang2018,Huang2019} This is principally due to their large absorption coefficient and the tunability of the bandgap of InGaN alloys over the whole visible spectrum. Moreover, their resistance to radiation and their high thermal stability make III-nitrides ideally suited for  photovoltaic applications.\cite{Williams2017,Huang2018,Huang2019} In particular, recent studies have shown good performance of InGaN/GaN MQW cells at high solar concentration and over a large temperature range.\cite{Moses2020}
The combination of high solar concentration and high temperature operation is of particular relevance for hybrid solar thermal-photovoltaic power conversion facilities.\cite{Zeitouny2018} 

The open-circuit voltage ($\Voc$) is of special importance for concentrated photovoltaics working at high temperatures.
While increasing irradiance increases $\Voc$, higher temperatures  decrease it.\cite{Braun2013}
The optimization of solar cell performance therefore requires a full understanding of the dependence of $\Voc$ on incident intensity and device temperature $T$.

$\Voc$ is related to the splitting of the electron and hole quasi-Fermi levels, which results from the balance of optical generation rate $G$ and recombination rate $R$, i.e. $R(n,p) = G$, where $n$ and $p$ are the respective carrier densities.
If a single recombination channel dominates and the Boltzmann approximation is valid, then assuming $n=p$ the difference of the quasi Fermi levels and thus $\Voc$ is given by\cite{Matthias_voc}
\begin{equation}
  \Voc = \frac{E_g}{q} + \frac{\eta k_BT}{q} \ln \frac{G}{C(N_cN_v)^\frac{1}{\eta}}. \label{eq:voc_bulk_rec}
\end{equation}
Here $E_g$ is the band gap energy, $k_B$ the Boltzmann constant, $q$ the elementary charge, $C$ the recombination parameter, and $N_{c,v}$ the electron and hole effective density of states. $G$ is assumed proportional to the incident light intensity.
The parameter $\eta$ is related to the degree of the recombination process and is 2, 1 or 2/3 for Shockley-Read-Hall, radiative and Auger recombination, respectively.\cite{Matthias_voc}

$\Voc$ is also described by the equivalent diode model\cite{Dupre_book,Braun2013} as  
\begin{equation}
  \Voc = \frac{E_g}{q} + \nktq \ln \frac{I_{\mathrm{ph}}}{I_0}, \label{eq:voc_diode} 
\end{equation}
where $I_0$ is the diode saturation current, $I_{ph}$ is the photo-generated current and $\eta$ is the diode ideality factor.
Comparing with \eqref{eq:voc_bulk_rec}, the diode ideality factor can be related to the dominant recombination process. Note that $\eta$ is bias-dependent since the dominant recombination process changes as carrier injection varies. It can be larger than 2 in the presence of other processes such as trap-assisted tunneling.\cite{AufDerMaur2014}

As seen from both models, the temperature and intensity dependence of $\Voc$ permit the determination of two important device parameters.
The extrapolation to $T =$ 0 K should provide $E_g$ independent of  $G$. The derivative of $\Voc$ with respect to $\ln G$ at fixed $T$ should yield a constant slope from which the value of $\eta$, and hence the dominant recombination mechanism can be deduced.
These two quantities can be obtained experimentally by temperature and intensity dependent measurements, respectively.

\begin{figure}[hbt]
\centering
  \includegraphics[width=\columnwidth]{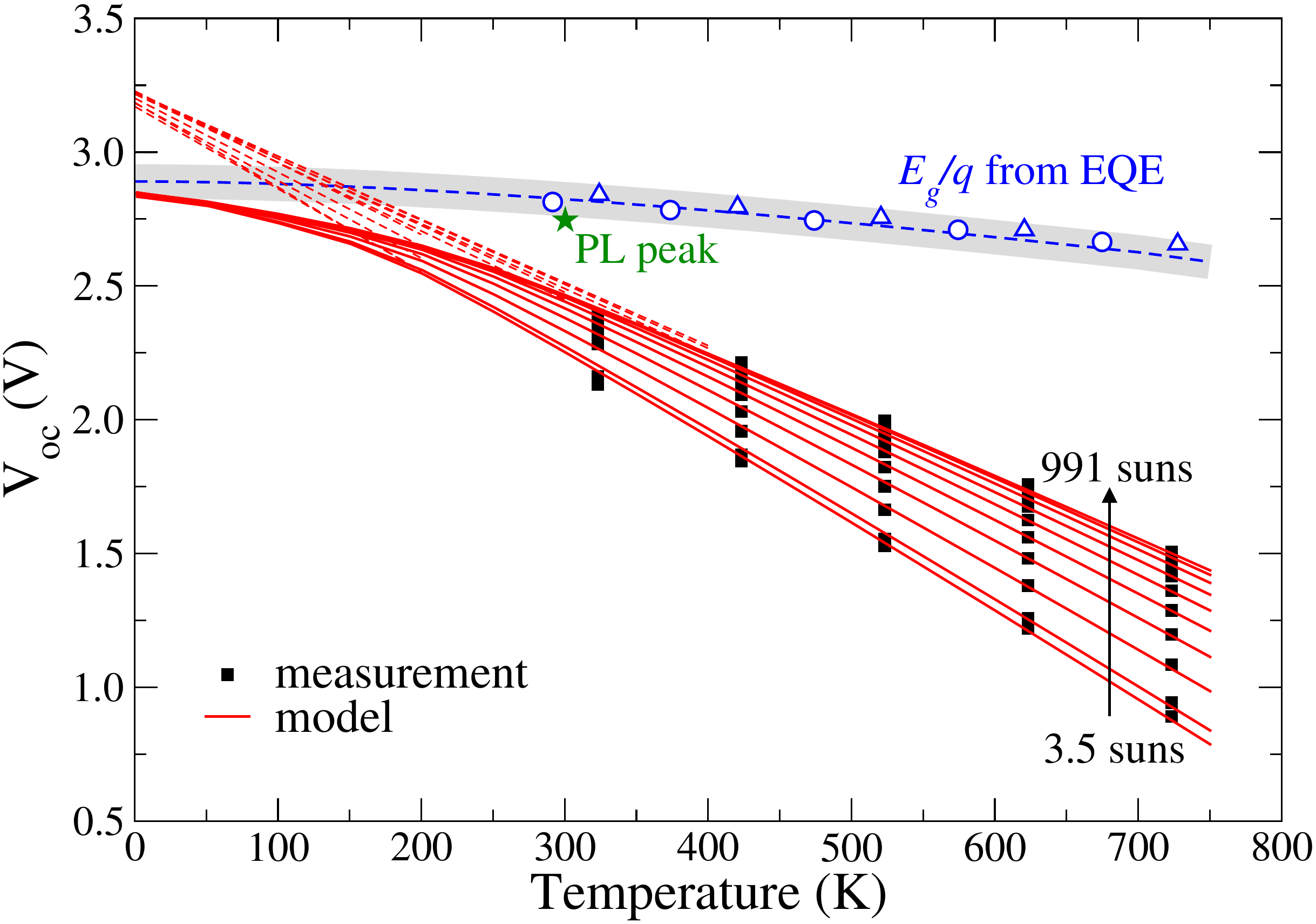}
  \caption{$\Voc$ versus temperature $T$ for the studied InGaN/GaN MQW solar cell.  The black squares are measured values for different intensities from 3.5 -- 991 suns.
  $E_g/q$ estimated from the EQE (blue open symbols) and PL (green star) data is indicated. The blue dashed line is a fit to the EQE data based on Varshni's law, with parameters from literature,\cite{vurgaftman} which allows an estimation of the lowest transition energy at 0 K. The grey shaded area is a guide to the eye, indicating the expected range of $E_g/q$. The experimental uncertainty in $\Voc$ and $E_g/q$ is on the order of mV and therefore not shown.}
\label{fig:voc_vs_temp}
\end{figure}

\begin{figure}[hbt]
\centering
  \includegraphics[width=\columnwidth]{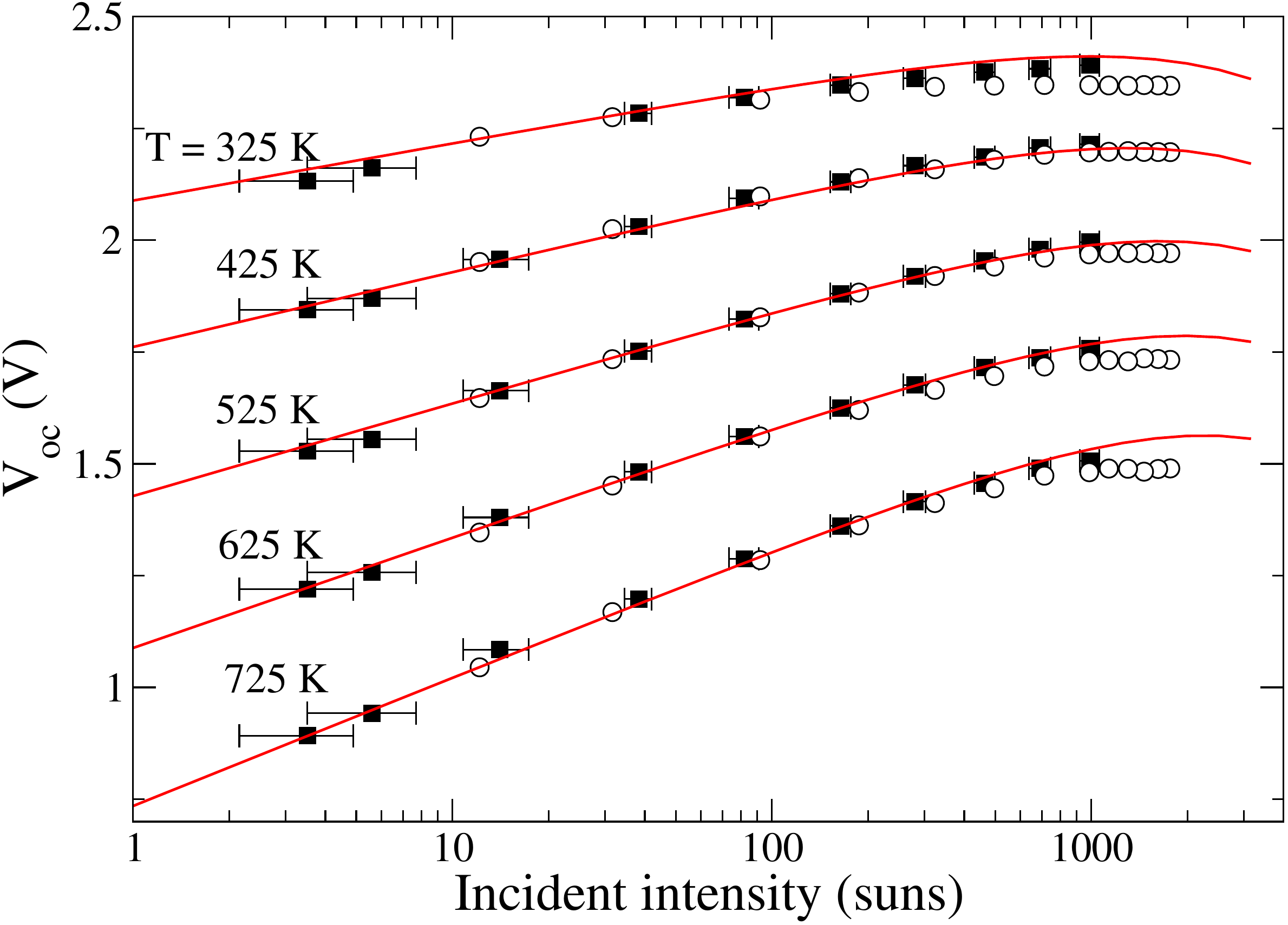}
  \caption{$\Voc$ versus solar intensity for the studied cell at different base plate temperatures. The solid black squares and open white circles are measured values from two different measurement series.
  The red lines are  model predictions as described in the text. The change in slope can be explained by self-heating of the cell under high intensity. Experimental uncertainty in  intensity is indicated for one of the data sets.
  }
\label{fig:voc_vs_intensity}
\end{figure}

Such experiments have been  performed recently on c-plane InGaN/GaN MQW solar cells,\cite{Moses2020} revealing discrepancies with the above deductions from the simple analytic models.
Specifically, the extrapolated value of $q\Voc$ was found to be larger than the $E_g$ extracted from photoluminescence (PL) and external quantum efficiency (EQE) measurements.
Moreover, at  intensities above $\sim 100$ suns, the $\Voc$ versus $\ln G$ curve changes slope.
Figure~\ref{fig:voc_vs_temp} shows the measured $\Voc$ as a function of $T$ for different intensities measured with a fiber-optic minidish solar concentrator.\cite{Katz2006,Gordon2004}
For the lower concentration regime, the  data show linear behavior as predicted from the model equations.
Linear extrapolations to 0 K at different intensities, indicated by dashed lines in the figure, indeed converge to a single value of $E_g \approx 3.15$ eV.
But this is $\sim 0.3$ eV higher than the $E_g$ extracted from both quantum efficiency and PL measurements, considering the temperature dependence of $E_g$ described by Varshni's law,\cite{vurgaftman} and also with respect to theoretical k$\cdot$p calculations we performed for this device structure.
Similarly, the slope $S=q/k_BT\cdot\partial\Voc/\partial\ln I$ of the $\Voc$ versus intensity data shown in Fig.~\ref{fig:voc_vs_intensity} is constant near a value of 2 up to $\sim 100$ suns, but then drops quickly to below 1. 
This suggests dominance of defect-related Shockley-Read-Hall (SRH) recombination at lower concentration, but also a transition to other recombination processes, or a thermal effect at higher intensity, or both.

We will now show that the data are compatible with the ABC model commonly used for the analysis of III-nitride light emitting diodes (LEDs),\cite{Karpov2015,David_2020} provided a temperature-dependent SRH parameter and device heating at high intensity are accounted for.
In the ABC model, $R$ is given as a sum of three contributions up to third order in the carrier density:
\begin{equation}
  R=An + Bn^2 + Cn^3.\label{eq:ABC}
\end{equation}
This model follows from standard recombination models\cite{Sze} under the assumption of equal electron and hole densities.
The first two terms are identified with SRH and radiative recombination, respectively, and the third with Auger recombination, although it might also model carrier leakage.\cite{Piprek2015}
It can be assumed that all recombination in the MQW structure is governed by the quantum well populations, so that $n$ is interpreted as the 2D electron density.
We also invoke the Boltzmann approximation, which is justified for the  intensities considered here. Furthermore, we identify $\Voc$ with the quasi-Fermi level splitting in the MQW region, and assume a homogeneous distribution of generation and recombination over the quantum wells. Then $\Voc$ can be obtained in closed form from \eqref{eq:ABC} equating $R=G$ and using formulas for solving cubic equations.\cite{nickalls_1993}
This leads to
\begin{subequations}
 \begin{equation}
    \Delta_0  = B^2 - 3AC,\; \Delta_1 = 2B^3 - 9ABC - 27 C^2G, \nonumber
 \end{equation}
 \begin{equation}\label{eq:cuberoot}
    \zeta  = \sqrt[3]{\frac{\Delta_1 \pm \sqrt{\Delta_1^2-4\Delta_0^3}}{2}}, 
 \end{equation}
 \begin{equation}
     n = -\frac{1}{3C}(B+\zeta+\frac{\Delta_0}{\zeta}),
 \end{equation}
 and finally
 \begin{equation}
     \Voc = \frac{E_g}{q}+\frac{2k_BT}{q}\ln\left[ -\frac{1}{3C\sqrt{N_cN_v}}(B+\zeta+\frac{\Delta_0}{\zeta})\right].
 \end{equation}
\end{subequations}
In \eqref{eq:cuberoot}, the physically meaningful root must insure that $n$ is real and positive. There is only one such root since $A, B, C, G > 0$.
For evaluating $E_g$, we used the measured PL peak energy and the extrapolation to 0 K via Varshni's law.
$G$ in each QW was estimated from the measured short-circuit current density $J_{sc}$ and EQE as $G=J_{sc}/(0.6N_{QW})$. $N_{QW} = 30$ is the number of QWs and 0.6 is a rough estimate for the extraction efficiency, which for simplicity has been taken as constant.
The 2D densities of states in the QWs are given by $N_{c,v} = {m_{e,h}k_BT}/{\pi\hbar^2}$, where we used for the effective masses $m_e$ and $m_h$ values from Ref. \onlinecite{vurgaftman}.
Table \ref{tab:params} lists all parameters and references.

At this point, it would be tempting to fit the parameters $A$, $B$ and $C$ to the data.
This, however, does not solve the problem of the overestimation of $E_g$ at $T=0$ K. In addition, it leads to unreasonably high values of $B$ and $C$ compared to their published values.\cite{Nippert2016,Zhou2020,David_2020}
Note, however, that a large $C$ parameter might be justified in the case of carrier leakage, although this is observed in LEDs only at high carrier injection.\cite{Piprek2015}
A large $B$ parameter, on the other hand, could be justified by supposing a trap-assisted Auger process, but signatures of this have so far been observed only in low-efficiency MBE-grown devices.\cite{Espenlaub2020}
Since values for $B$ and $C$ are relatively well established, both from measurements and theoretical predictions, we preferred to reduce the number of fitting parameters by using experimentally-extracted $B$ and $C$ from the work of Ref. \onlinecite{Nippert2016}.
Moreover, it is reasonable to assume that $B$ and $C$ have similar values in quantum wells of similar composition and thickness, as is the case here, while $A$ may be largely process dependent.

\begin{figure}
\centering
  \includegraphics[width=\columnwidth]{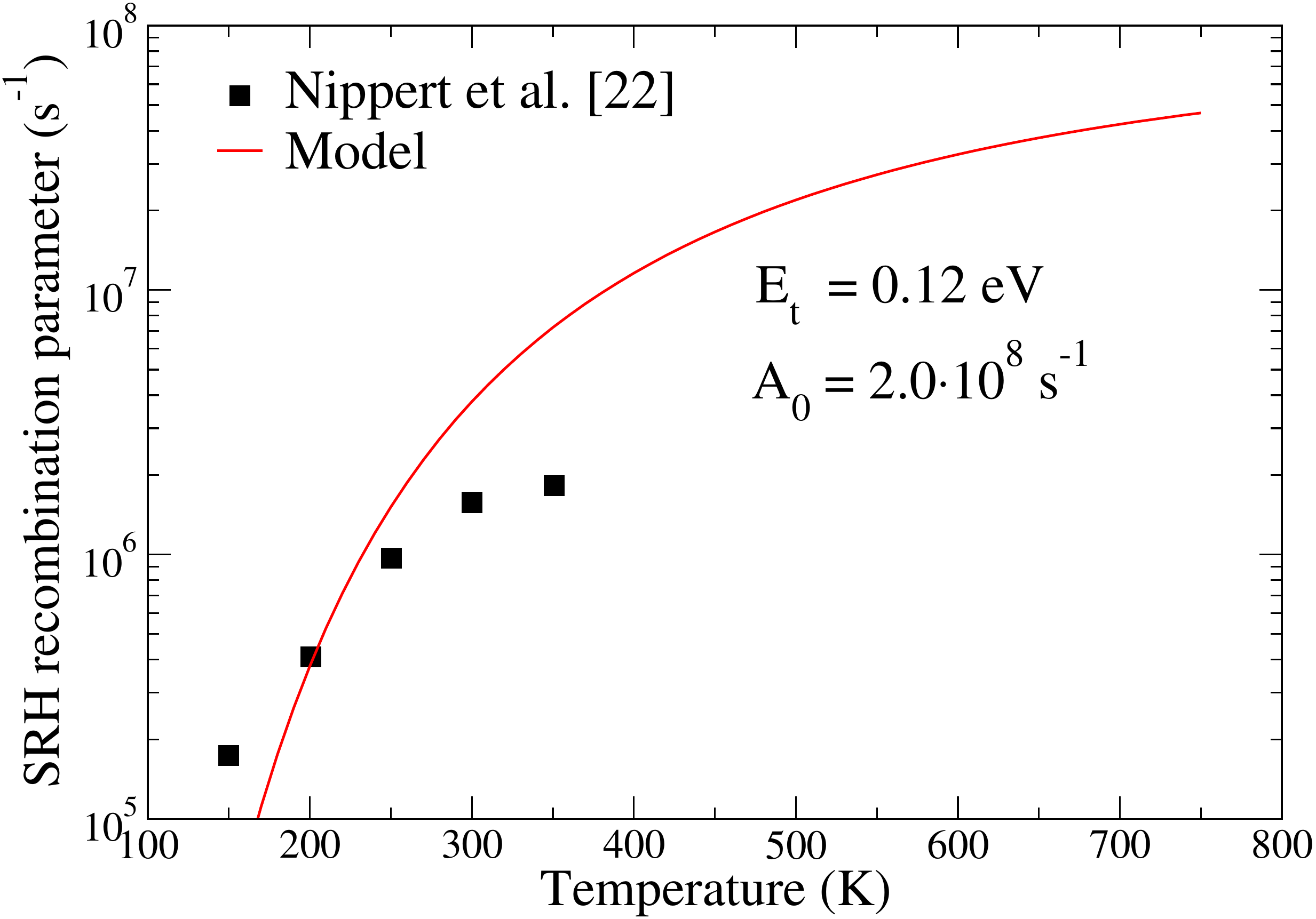}
  \caption{Temperature dependence of the SRH recombination parameter $A$ according to Eq. \eqref{eq:ABC}. The symbols show values from Ref. \onlinecite{Nippert2016}, where $A$ has been extracted from differential lifetime measurements (experimental uncertainty was not  provided). }
\label{fig:A_vs_T}
\end{figure}

In order to reproduce the observed thermal behavior, we adopted a temperature dependent model for $A$ of the form\cite{DeSanti2017,Rashidi2019}
\begin{equation}
  A(T)=\frac{A_0}{1+\cosh(E_T/k_BT)},
\end{equation}
where $A_0$ in $s^{-1}$ is a constant depending on defect density and properties of the defect state, and $E_T$ is the energy level of the defect state measured from the intrinsic Fermi level.
Figure~\ref{fig:A_vs_T} shows $A(T)$ in comparison with published values.\cite{Nippert2016}

Based on the measured short-circuit currents, it is reasonable to expect that at 1000 suns carrier injection is still roughly an order of magnitude below that at typical current densities at maximum internal quantum efficiency in InGaN/GaN LEDs.\cite{David_2020,Reklaitis_2018} Hence even at high solar concentration $\Voc$ is still largely limited by SRH recombination. Therefore, the measured change in slope of the $\Voc$ versus intensity curves apparently cannot be explained by the transition between dominant recombination mechanisms.

Instead, we posit a a non-negligible thermal effect, because under open circuit a major part of the absorbed optical power is transformed to heat
via thermalization and non-radiative recombination of photogenerated charge carriers.
The heat sink we used to maintain the cell at a constant base-plate temperature $T$ cannot remove this heat fast enough to prevent noticeable cell overheating once the cell irradiance exceeds several hundred suns.
To model this, we assumed a constant thermal resistance $R_\mathrm{Th}$ so that the device temperature is given by $T_\mathrm{MQW}=T+R_\mathrm{Th}P$, where $P$ is the absorbed power.
Using $P=0.1$ W/cm$^2$ (1 sun) and the active cell area of 0.125$\times$0.125 cm$^2$, we obtain $R_\mathrm{Th} \approx 17$ K/W.
This value is in the range that would be expected for our experimental configuration,\cite{Sun2005} and leads to an additional temperature increase of $\sim 27$ K at 1000 suns.

\begin{table}
  \centering
  \caption{Model parameters}\label{tab:params}
  \begin{tabular}{|l||c|l|c|}
    \hline
    \bf Parameter   & \bf Value & \bf Units & \bf Source\\
    \hline
    \hline
    $A_0$ & $2\cdot10^8$ & s$^{-1}$ & Fit \\
    \hline
    $E_T$ & 0.12 & eV & Fit \\
    \hline
    $B$ & $2\cdot10^{-5}$ & cm$^2$s$^{-1}$ & [\onlinecite{Nippert2016}] \\
    \hline
    $C$ & $5\cdot10^{-18}$ & cm$^4$s$^{-1}$ & [\onlinecite{Nippert2016}] \\
    \hline
    $E_g$       & 2.85 & eV & meas. \\
    \hline
    $m_e$ & 0.2 & - & [\onlinecite{vurgaftman}] \\
    \hline
    $m_h$ & 1.4 & - & [\onlinecite{vurgaftman}] \\
    \hline
    $R_{\mathrm{Th}}$ & 17 & K/W & Fit \\
    \hline
    \hline
  \end{tabular}
\end{table}

Figure \ref{fig:voc_vs_temp} presents the modeled $\Voc$ as a function of $T$ and a broad range of solar intensities up to $\sim 1000$ suns together with our measurements, using model parameters given in Table~\ref{tab:params}.
Data at higher intensities are not shown, because they are indistinguishable from the curve at 991 suns on the scale of the plot.
The linear regression of the measured data and the extrapolation to 0 K are given by the red dashed lines. 
Since the measured $q\Voc$ is expected to be smaller than $E_g$, or more specifically the ground-state transition energy, we compare it with the measured PL peak energy and $E_g$ extracted from EQE measurements. This is indicated in Fig. \ref{fig:voc_vs_temp} by the green star and the blue open symbols, respectively. In order to obtain $E_g$ at 0 K, we extrapolated the data according to Varshni's law for the temperature dependence of $E_g$, using published values for the model parameters and a linear interpolation between values for GaN and InN.\cite{vurgaftman}

The linear interpolation from the high-temperature $\Voc$ data leads to an overestimation of $E_g$ which is incompatible with the direct measurement.
Moreover, this precludes fitting the  data using the measured $E_g$. 
In contradistinction, it can be seen that the model employing a temperature-dependent SRH parameter $A$ predicts a change in slope at around 200 K, which is associated with a transition from SRH to radiative-dominated recombination. Setting $E_g$ in the analytic formulas to the value obtained by measurement, we can consistently reproduce the measurements using the correct $E_g$ at 0 K.
This shows that incorrect values of $E_g$ are obtained  from linear regression of the data around room temperature due to the temperature dependence of the recombination parameters.
Moreover, such a transition to a radiatively-dominated regime at low temperatures is compatible with the assumption often made for III-nitride LEDs that the PL efficiency approaches 100\% near 0 K.\cite{Watanabe2003}

Figure~\ref{fig:voc_vs_intensity} compares model results against the data as a function of  intensity.
The slope of 2 at low solar concentration is well reproduced, while the sub-linear behavior at larger concentration is accounted for by the additional heating as described above.
Note that the curves at different temperatures can be fitted conjointly thanks to the temperature dependent model for the SRH parameter. Trying to reproduce these curves by adjusting the recombination parameters $A$, $B$ and $C$ is possible, but leads to values of $B$ and $C$ which are up to 5 orders of magnitude larger than published values (vide supra).

In conclusion, we have shown that the measured intensity and temperature dependence of $\Voc$ of c-plane InGaN/GaN MQW solar cells is compatible with the ABC model describing recombination in such structures, provided a temperature-dependent SRH parameter is used and self-heating is taken into account.
The model can then correctly reproduce the high temperature behavior of $\Voc$, while recovering the correct value of $E_g$ at 0 K. The results also explain the discrepancy between the linear temperature extrapolation of  $\Voc$ to 0 K and the measured $E_g$.
This is important inasmuch as $\Voc$ at 0 K represents the maximum voltage that can be generated by a solar cell. In this context, it will be interesting to generalize our results to a wide range of solar cells based on direct and indirect semiconductors.

\begin{acknowledgments}

GM’s doctoral fellowship was funded by Research Grant number 3-15970 from the Israel Ministry of Science, Technology and Space, as well as by Ben-Gurion University's Albert Katz International School for Desert Studies and the Hitech Biotech Scholarship from the university's Kreitman School of Advanced Graduate Studies.

\end{acknowledgments}

\section*{Data availability}
The data that support the findings of this study are available from the corresponding authors upon reasonable request.

\bibliography{biblio}

\begin{thebibliography}{29}%
\makeatletter
\providecommand \@ifxundefined [1]{%
 \@ifx{#1\undefined}
}%
\providecommand \@ifnum [1]{%
 \ifnum #1\expandafter \@firstoftwo
 \else \expandafter \@secondoftwo
 \fi
}%
\providecommand \@ifx [1]{%
 \ifx #1\expandafter \@firstoftwo
 \else \expandafter \@secondoftwo
 \fi
}%
\providecommand \natexlab [1]{#1}%
\providecommand \enquote  [1]{``#1''}%
\providecommand \bibnamefont  [1]{#1}%
\providecommand \bibfnamefont [1]{#1}%
\providecommand \citenamefont [1]{#1}%
\providecommand \href@noop [0]{\@secondoftwo}%
\providecommand \href [0]{\begingroup \@sanitize@url \@href}%
\providecommand \@href[1]{\@@startlink{#1}\@@href}%
\providecommand \@@href[1]{\endgroup#1\@@endlink}%
\providecommand \@sanitize@url [0]{\catcode `\\12\catcode `\$12\catcode
  `\&12\catcode `\#12\catcode `\^12\catcode `\_12\catcode `\%12\relax}%
\providecommand \@@startlink[1]{}%
\providecommand \@@endlink[0]{}%
\providecommand \url  [0]{\begingroup\@sanitize@url \@url }%
\providecommand \@url [1]{\endgroup\@href {#1}{\urlprefix }}%
\providecommand \urlprefix  [0]{URL }%
\providecommand \Eprint [0]{\href }%
\providecommand \doibase [0]{http://dx.doi.org/}%
\providecommand \selectlanguage [0]{\@gobble}%
\providecommand \bibinfo  [0]{\@secondoftwo}%
\providecommand \bibfield  [0]{\@secondoftwo}%
\providecommand \translation [1]{[#1]}%
\providecommand \BibitemOpen [0]{}%
\providecommand \bibitemStop [0]{}%
\providecommand \bibitemNoStop [0]{.\EOS\space}%
\providecommand \EOS [0]{\spacefactor3000\relax}%
\providecommand \BibitemShut  [1]{\csname bibitem#1\endcsname}%
\let\auto@bib@innerbib\@empty
\bibitem [{\citenamefont {Dahal}\ \emph {et~al.}(2009)\citenamefont {Dahal},
  \citenamefont {Pantha}, \citenamefont {Li}, \citenamefont {Lin},\ and\
  \citenamefont {Jiang}}]{Dahal2009}%
  \BibitemOpen
  \bibfield  {author} {\bibinfo {author} {\bibfnamefont {R.}~\bibnamefont
  {Dahal}}, \bibinfo {author} {\bibfnamefont {B.}~\bibnamefont {Pantha}},
  \bibinfo {author} {\bibfnamefont {J.}~\bibnamefont {Li}}, \bibinfo {author}
  {\bibfnamefont {J.~Y.}\ \bibnamefont {Lin}}, \ and\ \bibinfo {author}
  {\bibfnamefont {H.~X.}\ \bibnamefont {Jiang}},\ }\href {\doibase
  10.1063/1.3081123} {\bibfield  {journal} {\bibinfo  {journal} {Applied
  Physics Letters}\ }\textbf {\bibinfo {volume} {94}},\ \bibinfo {pages}
  {063505} (\bibinfo {year} {2009})}\BibitemShut {NoStop}%
\bibitem [{\citenamefont {{Bhuiyan}}\ \emph {et~al.}(2012)\citenamefont
  {{Bhuiyan}}, \citenamefont {{Sugita}}, \citenamefont {{Hashimoto}},\ and\
  \citenamefont {{Yamamoto}}}]{Bhuiyan2012}%
  \BibitemOpen
  \bibfield  {author} {\bibinfo {author} {\bibfnamefont {A.~G.}\ \bibnamefont
  {{Bhuiyan}}}, \bibinfo {author} {\bibfnamefont {K.}~\bibnamefont {{Sugita}}},
  \bibinfo {author} {\bibfnamefont {A.}~\bibnamefont {{Hashimoto}}}, \ and\
  \bibinfo {author} {\bibfnamefont {A.}~\bibnamefont {{Yamamoto}}},\ }\href
  {\doibase 10.1109/JPHOTOV.2012.2193384} {\bibfield  {journal} {\bibinfo
  {journal} {IEEE Journal of Photovoltaics}\ }\textbf {\bibinfo {volume} {2}},\
  \bibinfo {pages} {276} (\bibinfo {year} {2012})}\BibitemShut {NoStop}%
\bibitem [{\citenamefont {{Williams}}\ \emph {et~al.}(2017)\citenamefont
  {{Williams}}, \citenamefont {{McFavilen}}, \citenamefont {{Fischer}},
  \citenamefont {{Ding}}, \citenamefont {{Young}}, \citenamefont {{Vadiee}},
  \citenamefont {{Ponce}}, \citenamefont {{Arena}}, \citenamefont
  {{Honsberg}},\ and\ \citenamefont {{Goodnick}}}]{Williams2017}%
  \BibitemOpen
  \bibfield  {author} {\bibinfo {author} {\bibfnamefont {J.~J.}\ \bibnamefont
  {{Williams}}}, \bibinfo {author} {\bibfnamefont {H.}~\bibnamefont
  {{McFavilen}}}, \bibinfo {author} {\bibfnamefont {A.~M.}\ \bibnamefont
  {{Fischer}}}, \bibinfo {author} {\bibfnamefont {D.}~\bibnamefont {{Ding}}},
  \bibinfo {author} {\bibfnamefont {S.}~\bibnamefont {{Young}}}, \bibinfo
  {author} {\bibfnamefont {E.}~\bibnamefont {{Vadiee}}}, \bibinfo {author}
  {\bibfnamefont {F.~A.}\ \bibnamefont {{Ponce}}}, \bibinfo {author}
  {\bibfnamefont {C.}~\bibnamefont {{Arena}}}, \bibinfo {author} {\bibfnamefont
  {C.~B.}\ \bibnamefont {{Honsberg}}}, \ and\ \bibinfo {author} {\bibfnamefont
  {S.~M.}\ \bibnamefont {{Goodnick}}},\ }\href {\doibase
  10.1109/JPHOTOV.2017.2756057} {\bibfield  {journal} {\bibinfo  {journal}
  {IEEE Journal of Photovoltaics}\ }\textbf {\bibinfo {volume} {7}},\ \bibinfo
  {pages} {1646} (\bibinfo {year} {2017})}\BibitemShut {NoStop}%
\bibitem [{\citenamefont {Park}\ \emph {et~al.}(2018)\citenamefont {Park},
  \citenamefont {Nandi}, \citenamefont {Sim}, \citenamefont {Um}, \citenamefont
  {Kang}, \citenamefont {Kim},\ and\ \citenamefont {Lee}}]{Park2018}%
  \BibitemOpen
  \bibfield  {author} {\bibinfo {author} {\bibfnamefont {J.-H.}\ \bibnamefont
  {Park}}, \bibinfo {author} {\bibfnamefont {R.}~\bibnamefont {Nandi}},
  \bibinfo {author} {\bibfnamefont {J.-K.}\ \bibnamefont {Sim}}, \bibinfo
  {author} {\bibfnamefont {D.-Y.}\ \bibnamefont {Um}}, \bibinfo {author}
  {\bibfnamefont {S.}~\bibnamefont {Kang}}, \bibinfo {author} {\bibfnamefont
  {J.-S.}\ \bibnamefont {Kim}}, \ and\ \bibinfo {author} {\bibfnamefont
  {C.-R.}\ \bibnamefont {Lee}},\ }\href {\doibase 10.1039/C8RA03127D}
  {\bibfield  {journal} {\bibinfo  {journal} {RSC Adv.}\ }\textbf {\bibinfo
  {volume} {8}},\ \bibinfo {pages} {20585} (\bibinfo {year}
  {2018})}\BibitemShut {NoStop}%
\bibitem [{\citenamefont {Bai}\ \emph {et~al.}(2018)\citenamefont {Bai},
  \citenamefont {Gong}, \citenamefont {Li}, \citenamefont {Zhang},\ and\
  \citenamefont {Wang}}]{Bai2018}%
  \BibitemOpen
  \bibfield  {author} {\bibinfo {author} {\bibfnamefont {J.}~\bibnamefont
  {Bai}}, \bibinfo {author} {\bibfnamefont {Y.}~\bibnamefont {Gong}}, \bibinfo
  {author} {\bibfnamefont {Z.}~\bibnamefont {Li}}, \bibinfo {author}
  {\bibfnamefont {Y.}~\bibnamefont {Zhang}}, \ and\ \bibinfo {author}
  {\bibfnamefont {T.}~\bibnamefont {Wang}},\ }\href {\doibase
  https://doi.org/10.1016/j.solmat.2017.10.005} {\bibfield  {journal} {\bibinfo
   {journal} {Solar Energy Materials and Solar Cells}\ }\textbf {\bibinfo
  {volume} {175}},\ \bibinfo {pages} {47} (\bibinfo {year} {2018})}\BibitemShut
  {NoStop}%
\bibitem [{\citenamefont {Huang}\ \emph {et~al.}(2018)\citenamefont {Huang},
  \citenamefont {Chen}, \citenamefont {Fu}, \citenamefont {Baranowski},
  \citenamefont {Montes}, \citenamefont {Yang}, \citenamefont {Fu},
  \citenamefont {Gunning}, \citenamefont {Koleske},\ and\ \citenamefont
  {Zhao}}]{Huang2018}%
  \BibitemOpen
  \bibfield  {author} {\bibinfo {author} {\bibfnamefont {X.}~\bibnamefont
  {Huang}}, \bibinfo {author} {\bibfnamefont {H.}~\bibnamefont {Chen}},
  \bibinfo {author} {\bibfnamefont {H.}~\bibnamefont {Fu}}, \bibinfo {author}
  {\bibfnamefont {I.}~\bibnamefont {Baranowski}}, \bibinfo {author}
  {\bibfnamefont {J.}~\bibnamefont {Montes}}, \bibinfo {author} {\bibfnamefont
  {T.-H.}\ \bibnamefont {Yang}}, \bibinfo {author} {\bibfnamefont
  {K.}~\bibnamefont {Fu}}, \bibinfo {author} {\bibfnamefont {B.~P.}\
  \bibnamefont {Gunning}}, \bibinfo {author} {\bibfnamefont {D.~D.}\
  \bibnamefont {Koleske}}, \ and\ \bibinfo {author} {\bibfnamefont
  {Y.}~\bibnamefont {Zhao}},\ }\href {\doibase 10.1063/1.5028530} {\bibfield
  {journal} {\bibinfo  {journal} {Applied Physics Letters}\ }\textbf {\bibinfo
  {volume} {113}},\ \bibinfo {pages} {043501} (\bibinfo {year}
  {2018})}\BibitemShut {NoStop}%
\bibitem [{\citenamefont {Huang}\ \emph {et~al.}(2019)\citenamefont {Huang},
  \citenamefont {Li}, \citenamefont {Fu}, \citenamefont {Li}, \citenamefont
  {Zhang}, \citenamefont {Chen}, \citenamefont {Fang}, \citenamefont {Fu},
  \citenamefont {DenBaars}, \citenamefont {Nakamura}, \citenamefont {Goodnick},
  \citenamefont {Ning}, \citenamefont {Fan},\ and\ \citenamefont
  {Zhao}}]{Huang2019}%
  \BibitemOpen
  \bibfield  {author} {\bibinfo {author} {\bibfnamefont {X.}~\bibnamefont
  {Huang}}, \bibinfo {author} {\bibfnamefont {W.}~\bibnamefont {Li}}, \bibinfo
  {author} {\bibfnamefont {H.}~\bibnamefont {Fu}}, \bibinfo {author}
  {\bibfnamefont {D.}~\bibnamefont {Li}}, \bibinfo {author} {\bibfnamefont
  {C.}~\bibnamefont {Zhang}}, \bibinfo {author} {\bibfnamefont
  {H.}~\bibnamefont {Chen}}, \bibinfo {author} {\bibfnamefont {Y.}~\bibnamefont
  {Fang}}, \bibinfo {author} {\bibfnamefont {K.}~\bibnamefont {Fu}}, \bibinfo
  {author} {\bibfnamefont {S.~P.}\ \bibnamefont {DenBaars}}, \bibinfo {author}
  {\bibfnamefont {S.}~\bibnamefont {Nakamura}}, \bibinfo {author}
  {\bibfnamefont {S.~M.}\ \bibnamefont {Goodnick}}, \bibinfo {author}
  {\bibfnamefont {C.-Z.}\ \bibnamefont {Ning}}, \bibinfo {author}
  {\bibfnamefont {S.}~\bibnamefont {Fan}}, \ and\ \bibinfo {author}
  {\bibfnamefont {Y.}~\bibnamefont {Zhao}},\ }\href {\doibase
  10.1021/acsphotonics.9b00655} {\bibfield  {journal} {\bibinfo  {journal} {ACS
  Photonics}\ }\textbf {\bibinfo {volume} {6}},\ \bibinfo {pages} {2096}
  (\bibinfo {year} {2019})}\BibitemShut {NoStop}%
\bibitem [{\citenamefont {Moses}\ \emph {et~al.}(2020)\citenamefont {Moses},
  \citenamefont {Huang}, \citenamefont {Zhao}, \citenamefont {Auf~der Maur},
  \citenamefont {Katz},\ and\ \citenamefont {Gordon}}]{Moses2020}%
  \BibitemOpen
  \bibfield  {author} {\bibinfo {author} {\bibfnamefont {G.}~\bibnamefont
  {Moses}}, \bibinfo {author} {\bibfnamefont {X.}~\bibnamefont {Huang}},
  \bibinfo {author} {\bibfnamefont {Y.}~\bibnamefont {Zhao}}, \bibinfo {author}
  {\bibfnamefont {M.}~\bibnamefont {Auf~der Maur}}, \bibinfo {author}
  {\bibfnamefont {E.~A.}\ \bibnamefont {Katz}}, \ and\ \bibinfo {author}
  {\bibfnamefont {J.~M.}\ \bibnamefont {Gordon}},\ }\href {\doibase
  https://doi.org/10.1002/pip.3326} {\bibfield  {journal} {\bibinfo  {journal}
  {Progress in Photovoltaics: Research and Applications}\ }\textbf {\bibinfo
  {volume} {28}},\ \bibinfo {pages} {1167} (\bibinfo {year}
  {2020})}\BibitemShut {NoStop}%
\bibitem [{\citenamefont {Zeitouny}\ \emph {et~al.}(2018)\citenamefont
  {Zeitouny}, \citenamefont {Lalau}, \citenamefont {Gordon}, \citenamefont
  {Katz}, \citenamefont {Flamant}, \citenamefont {Dollet},\ and\ \citenamefont
  {Vossier}}]{Zeitouny2018}%
  \BibitemOpen
  \bibfield  {author} {\bibinfo {author} {\bibfnamefont {J.}~\bibnamefont
  {Zeitouny}}, \bibinfo {author} {\bibfnamefont {N.}~\bibnamefont {Lalau}},
  \bibinfo {author} {\bibfnamefont {J.~M.}\ \bibnamefont {Gordon}}, \bibinfo
  {author} {\bibfnamefont {E.~A.}\ \bibnamefont {Katz}}, \bibinfo {author}
  {\bibfnamefont {G.}~\bibnamefont {Flamant}}, \bibinfo {author} {\bibfnamefont
  {A.}~\bibnamefont {Dollet}}, \ and\ \bibinfo {author} {\bibfnamefont
  {A.}~\bibnamefont {Vossier}},\ }\href {\doibase
  https://doi.org/10.1016/j.solmat.2018.03.004} {\bibfield  {journal} {\bibinfo
   {journal} {Solar Energy Materials and Solar Cells}\ }\textbf {\bibinfo
  {volume} {182}},\ \bibinfo {pages} {61} (\bibinfo {year} {2018})}\BibitemShut
  {NoStop}%
\bibitem [{\citenamefont {Braun}, \citenamefont {Katz},\ and\ \citenamefont
  {Gordon}(2013)}]{Braun2013}%
  \BibitemOpen
  \bibfield  {author} {\bibinfo {author} {\bibfnamefont {A.}~\bibnamefont
  {Braun}}, \bibinfo {author} {\bibfnamefont {E.~A.}\ \bibnamefont {Katz}}, \
  and\ \bibinfo {author} {\bibfnamefont {J.~M.}\ \bibnamefont {Gordon}},\
  }\href {https://onlinelibrary.wiley.com/doi/abs/10.1002/pip.2210} {\bibfield
  {journal} {\bibinfo  {journal} {Progress in Photovoltaics: Research and
  Applications}\ }\textbf {\bibinfo {volume} {21}},\ \bibinfo {pages} {1087}
  (\bibinfo {year} {2013})}\BibitemShut {NoStop}%
\bibitem [{\citenamefont {Auf~der Maur}\ and\ \citenamefont
  {Di~Carlo}(2019)}]{Matthias_voc}%
  \BibitemOpen
  \bibfield  {author} {\bibinfo {author} {\bibfnamefont {M.}~\bibnamefont
  {Auf~der Maur}}\ and\ \bibinfo {author} {\bibfnamefont {A.}~\bibnamefont
  {Di~Carlo}},\ }\href {\doibase https://doi.org/10.1016/j.solener.2019.05.056}
  {\bibfield  {journal} {\bibinfo  {journal} {Solar Energy}\ }\textbf {\bibinfo
  {volume} {187}},\ \bibinfo {pages} {358 } (\bibinfo {year}
  {2019})}\BibitemShut {NoStop}%
\bibitem [{\citenamefont {Dupré}, \citenamefont {Vaillon},\ and\ \citenamefont
  {Green}(2017)}]{Dupre_book}%
  \BibitemOpen
  \bibfield  {author} {\bibinfo {author} {\bibfnamefont {O.}~\bibnamefont
  {Dupré}}, \bibinfo {author} {\bibfnamefont {R.}~\bibnamefont {Vaillon}}, \
  and\ \bibinfo {author} {\bibfnamefont {M.~A.}\ \bibnamefont {Green}},\
  }\href@noop {} {\emph {\bibinfo {title} {Thermal Behavior of Photovoltaic
  Devices}}}\ (\bibinfo  {publisher} {Springer International Publishing AG},\
  \bibinfo {address} {Gewerbestrasse 11, 6330 Cham, Switzerland},\ \bibinfo
  {year} {2017})\BibitemShut {NoStop}%
\bibitem [{\citenamefont {Auf~der Maur}\ \emph {et~al.}(2014)\citenamefont
  {Auf~der Maur}, \citenamefont {Galler}, \citenamefont {Pietzonka},
  \citenamefont {Strassburg}, \citenamefont {Lugauer},\ and\ \citenamefont
  {Di~Carlo}}]{AufDerMaur2014}%
  \BibitemOpen
  \bibfield  {author} {\bibinfo {author} {\bibfnamefont {M.}~\bibnamefont
  {Auf~der Maur}}, \bibinfo {author} {\bibfnamefont {B.}~\bibnamefont
  {Galler}}, \bibinfo {author} {\bibfnamefont {I.}~\bibnamefont {Pietzonka}},
  \bibinfo {author} {\bibfnamefont {M.}~\bibnamefont {Strassburg}}, \bibinfo
  {author} {\bibfnamefont {H.}~\bibnamefont {Lugauer}}, \ and\ \bibinfo
  {author} {\bibfnamefont {A.}~\bibnamefont {Di~Carlo}},\ }\href {\doibase
  10.1063/1.4896970} {\bibfield  {journal} {\bibinfo  {journal} {Applied
  Physics Letters}\ }\textbf {\bibinfo {volume} {105}} (\bibinfo {year}
  {2014}),\ 10.1063/1.4896970}\BibitemShut {NoStop}%
\bibitem [{\citenamefont {Katz}, \citenamefont {Gordon},\ and\ \citenamefont
  {Feuermann}(2006)}]{Katz2006}%
  \BibitemOpen
  \bibfield  {author} {\bibinfo {author} {\bibfnamefont {E.~A.}\ \bibnamefont
  {Katz}}, \bibinfo {author} {\bibfnamefont {J.~M.}\ \bibnamefont {Gordon}}, \
  and\ \bibinfo {author} {\bibfnamefont {D.}~\bibnamefont {Feuermann}},\ }\href
  {\doibase https://doi.org/10.1002/pip.670} {\bibfield  {journal} {\bibinfo
  {journal} {Progress in Photovoltaics: Research and Applications}\ }\textbf
  {\bibinfo {volume} {14}},\ \bibinfo {pages} {297} (\bibinfo {year} {2006})},\
  \Eprint
  {http://arxiv.org/abs/https://onlinelibrary.wiley.com/doi/pdf/10.1002/pip.670}
  {https://onlinelibrary.wiley.com/doi/pdf/10.1002/pip.670} \BibitemShut
  {NoStop}%
\bibitem [{\citenamefont {Gordon}\ \emph {et~al.}(2004)\citenamefont {Gordon},
  \citenamefont {Katz}, \citenamefont {Feuermann},\ and\ \citenamefont
  {Huleihil}}]{Gordon2004}%
  \BibitemOpen
  \bibfield  {author} {\bibinfo {author} {\bibfnamefont {J.~M.}\ \bibnamefont
  {Gordon}}, \bibinfo {author} {\bibfnamefont {E.~A.}\ \bibnamefont {Katz}},
  \bibinfo {author} {\bibfnamefont {D.}~\bibnamefont {Feuermann}}, \ and\
  \bibinfo {author} {\bibfnamefont {M.}~\bibnamefont {Huleihil}},\ }\href
  {\doibase 10.1063/1.1723690} {\bibfield  {journal} {\bibinfo  {journal}
  {Applied Physics Letters}\ }\textbf {\bibinfo {volume} {84}},\ \bibinfo
  {pages} {3642} (\bibinfo {year} {2004})},\ \Eprint
  {http://arxiv.org/abs/https://doi.org/10.1063/1.1723690}
  {https://doi.org/10.1063/1.1723690} \BibitemShut {NoStop}%
\bibitem [{\citenamefont {Vurgaftman}, \citenamefont {Meyer},\ and\
  \citenamefont {Ram-Mohan}(2003)}]{vurgaftman}%
  \BibitemOpen
  \bibfield  {author} {\bibinfo {author} {\bibfnamefont {I.}~\bibnamefont
  {Vurgaftman}}, \bibinfo {author} {\bibfnamefont {J.}~\bibnamefont {Meyer}}, \
  and\ \bibinfo {author} {\bibfnamefont {L.}~\bibnamefont {Ram-Mohan}},\
  }\href@noop {} {\bibfield  {journal} {\bibinfo  {journal} {Applied Physics
  Review}\ }\textbf {\bibinfo {volume} {94}},\ \bibinfo {pages} {3675}
  (\bibinfo {year} {2003})}\BibitemShut {NoStop}%
\bibitem [{\citenamefont {Karpov}(2015)}]{Karpov2015}%
  \BibitemOpen
  \bibfield  {author} {\bibinfo {author} {\bibfnamefont {S.}~\bibnamefont
  {Karpov}},\ }\href {\doibase 10.1007/s11082-014-0042-9} {\bibfield  {journal}
  {\bibinfo  {journal} {Optical and Quantum Electronics}\ }\textbf {\bibinfo
  {volume} {47}},\ \bibinfo {pages} {1293} (\bibinfo {year}
  {2015})}\BibitemShut {NoStop}%
\bibitem [{\citenamefont {David}\ \emph {et~al.}(2020)\citenamefont {David},
  \citenamefont {Young}, \citenamefont {Lund},\ and\ \citenamefont
  {Craven}}]{David_2020}%
  \BibitemOpen
  \bibfield  {author} {\bibinfo {author} {\bibfnamefont {A.}~\bibnamefont
  {David}}, \bibinfo {author} {\bibfnamefont {N.~G.}\ \bibnamefont {Young}},
  \bibinfo {author} {\bibfnamefont {C.}~\bibnamefont {Lund}}, \ and\ \bibinfo
  {author} {\bibfnamefont {M.~D.}\ \bibnamefont {Craven}},\ }\href {\doibase
  10.1149/2.0372001jss} {\bibfield  {journal} {\bibinfo  {journal} {{ECS}
  Journal of Solid State Science and Technology}\ }\textbf {\bibinfo {volume}
  {9}},\ \bibinfo {pages} {016021} (\bibinfo {year} {2020})}\BibitemShut
  {NoStop}%
\bibitem [{\citenamefont {Sze}(1985)}]{Sze}%
  \BibitemOpen
  \bibfield  {author} {\bibinfo {author} {\bibfnamefont {S.~M.}\ \bibnamefont
  {Sze}},\ }\href@noop {} {\emph {\bibinfo {title} {{Semiconductor Devices:
  Physics and Technology}}}}\ (\bibinfo  {publisher} {John Wiley \& Sons, New
  York},\ \bibinfo {year} {1985})\BibitemShut {NoStop}%
\bibitem [{\citenamefont {Piprek}(2015)}]{Piprek2015}%
  \BibitemOpen
  \bibfield  {author} {\bibinfo {author} {\bibfnamefont {J.}~\bibnamefont
  {Piprek}},\ }\href {\doibase 10.1063/1.4927202} {\bibfield  {journal}
  {\bibinfo  {journal} {Applied Physics Letters}\ }\textbf {\bibinfo {volume}
  {107}},\ \bibinfo {pages} {031101} (\bibinfo {year} {2015})}\BibitemShut
  {NoStop}%
\bibitem [{\citenamefont {Nickalls}(1993)}]{nickalls_1993}%
  \BibitemOpen
  \bibfield  {author} {\bibinfo {author} {\bibfnamefont {R.}~\bibnamefont
  {Nickalls}},\ }\href {\doibase 10.2307/3619777} {\bibfield  {journal}
  {\bibinfo  {journal} {The Mathematical Gazette}\ }\textbf {\bibinfo {volume}
  {77}},\ \bibinfo {pages} {354–359} (\bibinfo {year} {1993})}\BibitemShut
  {NoStop}%
\bibitem [{\citenamefont {Nippert}\ \emph {et~al.}(2016)\citenamefont
  {Nippert}, \citenamefont {Karpov}, \citenamefont {Callsen}, \citenamefont
  {Galler}, \citenamefont {Kure}, \citenamefont {Nenstiel}, \citenamefont
  {Wagner}, \citenamefont {Straßburg}, \citenamefont {Lugauer},\ and\
  \citenamefont {Hoffmann}}]{Nippert2016}%
  \BibitemOpen
  \bibfield  {author} {\bibinfo {author} {\bibfnamefont {F.}~\bibnamefont
  {Nippert}}, \bibinfo {author} {\bibfnamefont {S.~Y.}\ \bibnamefont {Karpov}},
  \bibinfo {author} {\bibfnamefont {G.}~\bibnamefont {Callsen}}, \bibinfo
  {author} {\bibfnamefont {B.}~\bibnamefont {Galler}}, \bibinfo {author}
  {\bibfnamefont {T.}~\bibnamefont {Kure}}, \bibinfo {author} {\bibfnamefont
  {C.}~\bibnamefont {Nenstiel}}, \bibinfo {author} {\bibfnamefont {M.~R.}\
  \bibnamefont {Wagner}}, \bibinfo {author} {\bibfnamefont {M.}~\bibnamefont
  {Straßburg}}, \bibinfo {author} {\bibfnamefont {H.-J.}\ \bibnamefont
  {Lugauer}}, \ and\ \bibinfo {author} {\bibfnamefont {A.}~\bibnamefont
  {Hoffmann}},\ }\href {\doibase 10.1063/1.4965298} {\bibfield  {journal}
  {\bibinfo  {journal} {Applied Physics Letters}\ }\textbf {\bibinfo {volume}
  {109}},\ \bibinfo {pages} {161103} (\bibinfo {year} {2016})}\BibitemShut
  {NoStop}%
\bibitem [{\citenamefont {Zhou}\ \emph {et~al.}(2020)\citenamefont {Zhou},
  \citenamefont {Ikeda}, \citenamefont {Zhang}, \citenamefont {Liu},
  \citenamefont {Zhang}, \citenamefont {Tian}, \citenamefont {Wen},
  \citenamefont {Li}, \citenamefont {Zhang},\ and\ \citenamefont
  {Yang}}]{Zhou2020}%
  \BibitemOpen
  \bibfield  {author} {\bibinfo {author} {\bibfnamefont {R.}~\bibnamefont
  {Zhou}}, \bibinfo {author} {\bibfnamefont {M.}~\bibnamefont {Ikeda}},
  \bibinfo {author} {\bibfnamefont {F.}~\bibnamefont {Zhang}}, \bibinfo
  {author} {\bibfnamefont {J.}~\bibnamefont {Liu}}, \bibinfo {author}
  {\bibfnamefont {S.}~\bibnamefont {Zhang}}, \bibinfo {author} {\bibfnamefont
  {A.}~\bibnamefont {Tian}}, \bibinfo {author} {\bibfnamefont {P.}~\bibnamefont
  {Wen}}, \bibinfo {author} {\bibfnamefont {D.}~\bibnamefont {Li}}, \bibinfo
  {author} {\bibfnamefont {L.}~\bibnamefont {Zhang}}, \ and\ \bibinfo {author}
  {\bibfnamefont {H.}~\bibnamefont {Yang}},\ }\href {\doibase
  10.1063/1.5131716} {\bibfield  {journal} {\bibinfo  {journal} {Journal of
  Applied Physics}\ }\textbf {\bibinfo {volume} {127}},\ \bibinfo {pages}
  {013103} (\bibinfo {year} {2020})},\ \Eprint
  {http://arxiv.org/abs/https://doi.org/10.1063/1.5131716}
  {https://doi.org/10.1063/1.5131716} \BibitemShut {NoStop}%
\bibitem [{\citenamefont {Espenlaub}\ \emph {et~al.}(2019)\citenamefont
  {Espenlaub}, \citenamefont {Myers}, \citenamefont {Young}, \citenamefont
  {Marcinkevičius}, \citenamefont {Weisbuch},\ and\ \citenamefont
  {Speck}}]{Espenlaub2020}%
  \BibitemOpen
  \bibfield  {author} {\bibinfo {author} {\bibfnamefont {A.~C.}\ \bibnamefont
  {Espenlaub}}, \bibinfo {author} {\bibfnamefont {D.~J.}\ \bibnamefont
  {Myers}}, \bibinfo {author} {\bibfnamefont {E.~C.}\ \bibnamefont {Young}},
  \bibinfo {author} {\bibfnamefont {S.}~\bibnamefont {Marcinkevičius}},
  \bibinfo {author} {\bibfnamefont {C.}~\bibnamefont {Weisbuch}}, \ and\
  \bibinfo {author} {\bibfnamefont {J.~S.}\ \bibnamefont {Speck}},\ }\href
  {\doibase 10.1063/1.5096773} {\bibfield  {journal} {\bibinfo  {journal}
  {Journal of Applied Physics}\ }\textbf {\bibinfo {volume} {126}},\ \bibinfo
  {pages} {184502} (\bibinfo {year} {2019})}\BibitemShut {NoStop}%
\bibitem [{\citenamefont {Santi}\ \emph {et~al.}(2017)\citenamefont {Santi},
  \citenamefont {Meneghini}, \citenamefont {Monti}, \citenamefont {Glaab},
  \citenamefont {Guttmann}, \citenamefont {Rass}, \citenamefont {Einfeldt},
  \citenamefont {Mehnke}, \citenamefont {Enslin}, \citenamefont {Wernicke},
  \citenamefont {Kneissl}, \citenamefont {Meneghesso},\ and\ \citenamefont
  {Zanoni}}]{DeSanti2017}%
  \BibitemOpen
  \bibfield  {author} {\bibinfo {author} {\bibfnamefont {C.~D.}\ \bibnamefont
  {Santi}}, \bibinfo {author} {\bibfnamefont {M.}~\bibnamefont {Meneghini}},
  \bibinfo {author} {\bibfnamefont {D.}~\bibnamefont {Monti}}, \bibinfo
  {author} {\bibfnamefont {J.}~\bibnamefont {Glaab}}, \bibinfo {author}
  {\bibfnamefont {M.}~\bibnamefont {Guttmann}}, \bibinfo {author}
  {\bibfnamefont {J.}~\bibnamefont {Rass}}, \bibinfo {author} {\bibfnamefont
  {S.}~\bibnamefont {Einfeldt}}, \bibinfo {author} {\bibfnamefont
  {F.}~\bibnamefont {Mehnke}}, \bibinfo {author} {\bibfnamefont
  {J.}~\bibnamefont {Enslin}}, \bibinfo {author} {\bibfnamefont
  {T.}~\bibnamefont {Wernicke}}, \bibinfo {author} {\bibfnamefont
  {M.}~\bibnamefont {Kneissl}}, \bibinfo {author} {\bibfnamefont
  {G.}~\bibnamefont {Meneghesso}}, \ and\ \bibinfo {author} {\bibfnamefont
  {E.}~\bibnamefont {Zanoni}},\ }\href {\doibase 10.1364/PRJ.5.000A44}
  {\bibfield  {journal} {\bibinfo  {journal} {Photon. Res.}\ }\textbf {\bibinfo
  {volume} {5}},\ \bibinfo {pages} {A44} (\bibinfo {year} {2017})}\BibitemShut
  {NoStop}%
\bibitem [{\citenamefont {Rashidi}\ \emph {et~al.}(2019)\citenamefont
  {Rashidi}, \citenamefont {Monavarian}, \citenamefont {Aragon},\ and\
  \citenamefont {Feezell}}]{Rashidi2019}%
  \BibitemOpen
  \bibfield  {author} {\bibinfo {author} {\bibfnamefont {A.}~\bibnamefont
  {Rashidi}}, \bibinfo {author} {\bibfnamefont {M.}~\bibnamefont {Monavarian}},
  \bibinfo {author} {\bibfnamefont {A.}~\bibnamefont {Aragon}}, \ and\ \bibinfo
  {author} {\bibfnamefont {D.}~\bibnamefont {Feezell}},\ }\href {\doibase
  10.1038/s41598-019-56390-2} {\bibfield  {journal} {\bibinfo  {journal}
  {Scientific Reports}\ }\textbf {\bibinfo {volume} {9}} (\bibinfo {year}
  {2019}),\ 10.1038/s41598-019-56390-2}\BibitemShut {NoStop}%
\bibitem [{\citenamefont {Reklaitis}\ \emph {et~al.}(2018)\citenamefont
  {Reklaitis}, \citenamefont {Krencius}, \citenamefont {Malinauskas},
  \citenamefont {Karpov}, \citenamefont {Lugauer}, \citenamefont {Pietzonka},
  \citenamefont {Strassburg}, \citenamefont {Vitta},\ and\ \citenamefont
  {Toma{\v{s}}i{\={u}}nas}}]{Reklaitis_2018}%
  \BibitemOpen
  \bibfield  {author} {\bibinfo {author} {\bibfnamefont {I.}~\bibnamefont
  {Reklaitis}}, \bibinfo {author} {\bibfnamefont {L.}~\bibnamefont {Krencius}},
  \bibinfo {author} {\bibfnamefont {T.}~\bibnamefont {Malinauskas}}, \bibinfo
  {author} {\bibfnamefont {S.~Y.}\ \bibnamefont {Karpov}}, \bibinfo {author}
  {\bibfnamefont {H.~J.}\ \bibnamefont {Lugauer}}, \bibinfo {author}
  {\bibfnamefont {I.}~\bibnamefont {Pietzonka}}, \bibinfo {author}
  {\bibfnamefont {M.}~\bibnamefont {Strassburg}}, \bibinfo {author}
  {\bibfnamefont {P.}~\bibnamefont {Vitta}}, \ and\ \bibinfo {author}
  {\bibfnamefont {R.}~\bibnamefont {Toma{\v{s}}i{\={u}}nas}},\ }\href {\doibase
  10.1088/1361-6641/aaef06} {\bibfield  {journal} {\bibinfo  {journal}
  {Semiconductor Science and Technology}\ }\textbf {\bibinfo {volume} {34}},\
  \bibinfo {pages} {015007} (\bibinfo {year} {2018})}\BibitemShut {NoStop}%
\bibitem [{\citenamefont {Sun}\ \emph {et~al.}(2005)\citenamefont {Sun},
  \citenamefont {Israeli}, \citenamefont {Reddy}, \citenamefont {Scoles},
  \citenamefont {Gordon},\ and\ \citenamefont {Feuermann}}]{Sun2005}%
  \BibitemOpen
  \bibfield  {author} {\bibinfo {author} {\bibfnamefont {J.}~\bibnamefont
  {Sun}}, \bibinfo {author} {\bibfnamefont {T.}~\bibnamefont {Israeli}},
  \bibinfo {author} {\bibfnamefont {T.~A.}\ \bibnamefont {Reddy}}, \bibinfo
  {author} {\bibfnamefont {K.}~\bibnamefont {Scoles}}, \bibinfo {author}
  {\bibfnamefont {J.~M.}\ \bibnamefont {Gordon}}, \ and\ \bibinfo {author}
  {\bibfnamefont {D.}~\bibnamefont {Feuermann}},\ }\href {\doibase
  10.1115/1.1785799} {\bibfield  {journal} {\bibinfo  {journal} {Journal of
  Solar Energy Engineering}\ }\textbf {\bibinfo {volume} {127}},\ \bibinfo
  {pages} {138} (\bibinfo {year} {2005})}\BibitemShut {NoStop}%
\bibitem [{\citenamefont {Watanabe}\ \emph {et~al.}(2003)\citenamefont
  {Watanabe}, \citenamefont {Yamada}, \citenamefont {Nagashima}, \citenamefont
  {Ueki}, \citenamefont {Sasaki}, \citenamefont {Yamada}, \citenamefont
  {Taguchi}, \citenamefont {Tadatomo}, \citenamefont {Okagawa},\ and\
  \citenamefont {Kudo}}]{Watanabe2003}%
  \BibitemOpen
  \bibfield  {author} {\bibinfo {author} {\bibfnamefont {S.}~\bibnamefont
  {Watanabe}}, \bibinfo {author} {\bibfnamefont {N.}~\bibnamefont {Yamada}},
  \bibinfo {author} {\bibfnamefont {M.}~\bibnamefont {Nagashima}}, \bibinfo
  {author} {\bibfnamefont {Y.}~\bibnamefont {Ueki}}, \bibinfo {author}
  {\bibfnamefont {C.}~\bibnamefont {Sasaki}}, \bibinfo {author} {\bibfnamefont
  {Y.}~\bibnamefont {Yamada}}, \bibinfo {author} {\bibfnamefont
  {T.}~\bibnamefont {Taguchi}}, \bibinfo {author} {\bibfnamefont
  {K.}~\bibnamefont {Tadatomo}}, \bibinfo {author} {\bibfnamefont
  {H.}~\bibnamefont {Okagawa}}, \ and\ \bibinfo {author} {\bibfnamefont
  {H.}~\bibnamefont {Kudo}},\ }\href {\doibase 10.1063/1.1633672} {\bibfield
  {journal} {\bibinfo  {journal} {Applied Physics Letters}\ }\textbf {\bibinfo
  {volume} {83}},\ \bibinfo {pages} {4906} (\bibinfo {year}
  {2003})}\BibitemShut {NoStop}%
\end{thebibliography}%

\end{document}